\documentstyle[12pt]{article}

\begin{document}
\title{\textbf{Dynamos driven by poloidal flows in untwisted, curved and flat Riemannian diffusive flux tubes}}
\maketitle
{\sl \textbf{L.C. Garcia de Andrade}\newline
Departamento de F\'{\i}sica
Te\'orica-IF-UERJ- RJ, Brasil\\[-3mm]
\vspace{0.1cm} \paragraph*{Recently Vishik anti-fast dynamo theorem , has been tested against non-stretching flux
tubes [Phys Plasmas 15 (2008)]. In this paper, another anti-dynamo theorem, called Cowling's theorem, which states that axisymmetric magnetic fields cannot support dynamo action,
is carefully tested against thick tubular and curved Riemannian untwisted flows, as well as thin
flux tubes in diffusive and diffusionless media. In the non-diffusive media the Cowling's theorem
is not violated in thin Riemann-flat untwisted flux tubes, where the Frenet curvature is negative.
Nevertheless the diffusion action in the thin flux tube leads to a a dynamo action
driven by poloidal flows as shown by Love and Gubbins (Geophysical Res.) in the
context of geodynamos. Actually it is shown that a slow dynamo action is obtained.
In this case the Frenet and Riemann curvature still vanishes. In the case of magnetic
filaments in diffusive media dynamo action is obtained when the Frenet scalar curvature is negative. Since the Riemann curvature tensor can be expressed in terms of
the Frenet curvature of the magnetic flux tube axis, this result can be analogous to a
recent result obtained by Chicone, Latushkin and Smith, which states that geodesic
curvature in compact Riemannian manifolds can drive dynamo action in the manifold. It is also shown that in absence of diffusion, magnetic energy does not grow
but magnetic toroidal magnetic field can be generated by the poloidal field, what
is called a plasma dynamo.Key-words:anti-dynamo theorems, Cowling's theorem. PACS
numbers: 2.40.Hw: differential geometries. 91.25.Cw-dynamo theories.}
\newpage
\section{Introduction}
The most famous anti-dynamo theorems, considered by Zeldovich \cite{1}, and Cowling \cite{2}, have
told us how to recognise dynamo action in the most various types of flows and situations. In
the stretch-twist and fold (STF) Vainshtein-Zeldovich \cite{3,4} technique, besides the fundamental role played by stretching in dynamo action, another topological concept, called twist \cite{5}
comes into play. Along with stretch and twist, folding, which is associated with Riemann curvature \cite{6} may provide the doubling of the magnetic field which guarantees
dynamo action. Another very recent way to bypass Cowling's theorem in axisymmetric flows was recently investigated by Gissinger et al \cite{7} and consists of avoiding
perturbing the flow non-axisymmetrically. Therefore, is not always clear that the
simple lack of the twist ingredient may necessarily leads to a non-dynamo. Actually, untwisting tubes may give rise to an axisymmetric magnetic topology which by
Cowling's anti-dynamo theorem \cite{6}, would per se, guarantees the failure of dynamo
action. Actually, in the best of hypothesis, this left us with a slow dynamo, which
can be detected by performing the non-diffusive medium limit. The presence of diffusion is fundamental, in order to able one to perform the folding stage of the STF
dynamo mechanism. The dynamo thickness, maybe understood as proportional to
$Rm^{-1}$, which is the magnetic Reynolds number. This means that, large-scale astrophysical dynamos, where Rm acquire very large values, can be considered as thin
dynamos. On the other hand, small-scale dynamos such as the Perm toroidal experimental \cite{8} dynamo , can be considered as a thick dynamo. Besides applications in
dynamo theory, reconnection of untwisting flux tube gives rise to coalescing twisted
magnetic flux tubes. This was shown by Linton and Priest \cite{9} using a numerical
simulation. Another interesting anti-fast dynamo has been recently investigated by
Klapper and Young \cite{10}, who have extended the C1-differentiable anti-fast-dynamo
theorem by Vishik \cite{11} to C2. Previously, Arnold, Zeldovich, Ruzmaikin and Sokoloff
\cite{12} and Chicone and Latushkin \cite{13} have shown that uniform stretching in flows
could lead to fast dynamos in compact Riemannian manifolds. Yet more recently,
Garcia de Andrade \cite{14} have presented a new class of conformal Riemannian dynamos in three-dimensional diffusion substrate. In this paper the the zero-resistivity
ideal plasma limit in kinematic dynamos, is used to show that the thin untwisted
tubes cannot support dynamo action in accordance with Cowling's theorem. Recent
investigations by Nunez \cite{15} and Brandenburg \cite{16} have investigated the testing of
Cowling's theorem in Einsteins general relativistic black hole accretion disks \cite{17}.
Untwisting magnetic flux tube has been also investigated recently in the context of
solar physics by Terradas et al \cite{18}. They basically show that, untwisted flux tubes
possess a kink instability by shearing motion in diffusive medium. This situation,
often takes place in the solar surface, and in conformal fast dynamos in diffusive
media. Diffusion processes in Riemannian manifolds have been already considered
by S. Molchanov \cite{19}, where however, the problem of dynamo action was not addressed. Earlier Soward \cite{20}, argued that fast dynamo actions would still be possible
in regions where no non-stretching flows would be presented, such as in some curved
surfaces. Here one takes advantage of Soward's argument to build these surfaces
as Riemannian surfaces of arbitrarily curvature. The only constraint is that the flux
tubes remains in principle untwisted, and that the curvature be strong in diffusive
media, and of course that the magnetic flux tube axis be planar. To resume it seems
that Cowling's anti-dynamo theorem is not violated when the Riemann curvature
of the flux tube vanishes, a condition that mathematicians and general relativists
call Riemann-flat. The paper is organized as follows: In section II the the Cowling's
theorem is tested in Riemann-flat thin tubes in diffusive and non-dissipative media.
In section III the Cowling's theorem is shown to be violated when the Riemann
curvature is turned in diffusive media. In section IV diffusive fiments are shown
to drive dynamo action when normal stretching is considered. A similar result has
been obtained by Love and Gubbins \cite{21} in the case of geodynamos. Discussions
and future prospects are presented in section V.
\section{Cowling theorem in thin untwisted Riemann-flat magnetic flux tubes}Since we are dealing with Cowling theorem, we repeat here, for reader convenience,
the Cowling theorem. This theorem states that, dynamo action cannot be supported
in systems with axisymmetric flows and magnetic fields. In this section, a thin
untwisted Riemann-flat flux tube is shown to present a magnetic field decay or
at best a slow dynamo action, showing that Cowling's theorem is not violated in
these substrate. In this case one shall drop the Cartesian coordinates used here
and consider the Riemannian tubular coordinates, in principle curved Riemannian
flux tube. In the next section we shall observe that the absence of twisting only
can be substitute by strong curvature (folding) and stretching in the metric, with
the aid of a expressive thickness in the tube in diffusionless case. Actually in
the best hypothesis a slow dynamo developes. To start with let us consider the
Riemannian geometry of flux tubes as considered by Ricca \cite{22} and magnetic flux
tube coordinates $(r,{\theta}_{R},s)$, yields, which is also used in plasma torus called tokamaks,
by plasma physicists. Actually since the folding in flux tubes maybe be represented
by the Riemann curvature tensor, destructive folding that leads to fast dynamos,
can be obtained by the vanishing of folding or by vanishing of the Riemann curvature
tensor. The general flux tube Riemannian metric is
\begin{equation}
{{ds}_{0}}^{2}= dr^{2} + r^{2}d{\theta}^{2} + K^{2}(r,s)ds^{2}
\label{1}
\end{equation}
The thin Riemann-flat in principle twisted magnetic flux tube metric is obtained
by making the $K^{2}:=(1-r{\kappa}(s)cos{\theta})$ equal to one. This is clearly obtained when
the coordinate r approaches zero. This situation happens when one approaches the
torsioned flux tube axis. Here coordinate ${\theta}(s)$ is one of the Riemannian curvilinear
coordinates $(r,{\theta}_{R},s)$ and ${\theta}(s)={\theta}_{R}-\int{\tau(s)ds}$, ${\tau}$ being the Frenet torsion. Note that the thin tube metric is
\begin{equation}
{{ds}_{0}}^{2}= dr^{2} + r^{2}d{\theta}^{2} + ds^{2}
\label{2}
\end{equation}
Here the torsion term is responsible for the twist of the tube. Since
the tube is untwisted we assume here that the torsion vanishes and coordinate
${\theta}(s)={\theta}_{R}$. In general even the solar flux tubes are closed in the inner parts of the Sun, in the background of this compact Riemannian manifold, the Riemannian gradient
compact operator is given in general diffusive media by
\begin{equation}
\nabla={\textbf{e}_{r}}{\partial}_{r}+{\textbf{e}_{\theta}}\frac{1}{r}{\partial}_{\theta}+\textbf{t}\frac{1}{K}{\partial}_{s}
\label{3}
\end{equation}
\begin{equation}
d_{t}{\textbf{B}}={\eta}{\Delta}\textbf{B}+{\textbf{B}}.\nabla\textbf{v}
\label{4}
\end{equation}
where ${\Delta}={\nabla}^{2}$ is the Laplacian in general curvilinear coordinates and $\eta=0$ in this section, is the plasma resistivity or dissipation coefficient.
By considering that the magnetic field is strictly confined along and inside the tube,
and that the resistivity free case leads to the frozen condition, the flow is also
confined on the tube which allows us not to stretch the magnetic flux tube without
stretching the magnetic field. Since the tube is untwisted the following relation is simplified
\begin{equation}
{\partial}_{s}{\textbf{e}_{\theta}}=-{\tau}sin{\theta}\textbf{t}
\label{5}
\end{equation}
since the torsion vanishes
\begin{equation}
{\partial}_{\theta}{\textbf{e}_{r}}={\textbf{e}_{\theta}}
\label{6}
\end{equation}
and
\begin{equation}
{\partial}_{r}{\textbf{e}_{\theta}}=-\textbf{e}_{r}
\label{7}
\end{equation}
Together with the expression
\begin{equation}
{\partial}_{t}{\textbf{B}}={\gamma}{\textbf{B}}-{\omega}_{0}{B}_{\theta}\textbf{e}_{\theta}
\label{8}
\end{equation}
whose extra term is a non-inertial term similar to one that is introduced into a inertial
frame by the use of curvilinear coordinates, (Coriolis force in the frame) along with
two last equations helps to simplify analytical computations on the self-induction
equation, which can be splitted into the equation for the general untwisted tubes
\begin{equation}
{\textbf{B}}.{\nabla}\textbf{v}=\frac{1}{K}[{\partial}_{\theta}{\textbf{e}}_{\theta}+v_{s}{\kappa}{\textbf{n}}_{s}]B_{s}
+\frac{1}{r}{B}_{\theta}{\partial}_{\theta}{\textbf{e}}_{\theta}
\label{9}
\end{equation}
which applied to the thin tubes along with expression (\ref{8}) yields the following
self-induction scalar components along the Frenet frame $(\textbf{t},\textbf{n},\textbf{b})$, yields
\begin{equation}
-{\gamma}{B}_{\theta}sin{\theta}-{\omega}_{0}{B}_{\theta}cos{\theta}={B}_{s}v_{s}{\kappa}- {\omega}_{0}{B}_{\theta}cos{\theta}
\label{10}
\end{equation}
\begin{equation}
-{\gamma}{B}_{\theta}cos{\theta}-{\omega}_{0}{B}_{\theta}sin{\theta}=-{\omega}_{0}{B}_{\theta}sin{\theta}
\label{11}
\end{equation}
\begin{equation}
{\gamma}{B}_{s}=0
\label{12}
\end{equation}
Thus the last equation shows immeadiatly that no dynamo action can be supported
in this thin tube in diffusionless plasma medium. The remaining equations show that
the Frenet curvature ${\kappa}(s)$ vanishes. Note that from equation (\ref{8}) one obtains
\begin{equation}
{B}_{s}v_{s}{\kappa}=0
\label{13}
\end{equation}
This shows that the tube is untwisted, or better torsion-free, since twisting possesses
a non-torsion contribution, the toroidal flow does not necessarily vanishes, but can
vanish as well, which is the result given by Love and Gubbins  in the context
of geodynamos. Love and Gubbins have obtained this result numerically and not
analytically as here. Since, as one shall see in the next section, the Riemann curvature is
related to the Frenet curvature, which allows us to say that the tube is Riemann-flat. The
solenoidal condition
\begin{equation}
div{\textbf{B}}=0
\label{14}
\end{equation}
implies that
\begin{equation}
{\partial}_{s}{B}_{\theta}={\kappa}{\tau}rsin{\theta}{B}_{\theta}
\label{15}
\end{equation}
which in the absence of torsion shows that the poloidal component of the magnetic field does
not depend on the the toroidal coordinate , or
\begin{equation}
{\partial}_{s}{B}_{\theta}=0
\label{16}
\end{equation}
To simplify matters, this expression has been also used in the above computations.
One has also used the same equation for the incompressible plasma flow to obtain
\begin{equation}
{\partial}_{s}{v}_{\theta}=0
\label{17}
\end{equation}
which allows us to define the vorticity ${\omega}_{0}$ as
\begin{equation}
{v}_{\theta}={\omega}_{0}r
\label{18}
\end{equation}
This also simplify the above expressions. Other set of equations used in the computations are,
the Frenet steady equations
\begin{equation}
\frac{d\textbf{t}}{ds}={\kappa}(s)\textbf{n}
\label{19}
\end{equation}
\begin{equation}
\frac{d\textbf{n}}{ds}=-{\kappa}(s)\textbf{t}+{\tau}\textbf{b}
\label{20}
\end{equation}
\begin{equation}
\frac{d\textbf{b}}{ds}=-{\tau}(s)\textbf{n}
\label{21}
\end{equation}
Note that, to introduce diffusion now, the Laplacian term is needed. This can be written as
\begin{equation}
{\Delta}{\textbf{B}}=[{\textbf{t}}[{{\partial}_{r}}^{2}+\frac{1}{r}{{\partial}_{r}}-{\kappa}^{2}]B_{s}+
{\textbf{e}_{\theta}}[{{\partial}_{r}}^{2}+\frac{1}{r}{{\partial}_{r}}-\frac{1}{r^{2}}]B_{\theta}+
+(\frac{d}{ds}){\kappa}B_{s}{\textbf{n}}]
\label{22}
\end{equation}
Thus, for thin tubes the above Riemannian Laplacian operators, inside brackets is substituted
into the self-induction equation yielding the following differential equations in the approxima-
tion of linear Frenet curvature, yields
\begin{equation}
[{{\partial}_{r}}^{2}+\frac{1}{r}{{\partial}_{r}}-\frac{\gamma}{\eta}]B_{s}=0
\label{23}
\end{equation}
\begin{equation}
[{{\partial}_{r}}^{2}+\frac{1}{r}{{\partial}_{r}}-[\frac{\gamma}{\eta}+\frac{1}{r^{2}}]]B_{\theta}=0
\label{24}
\end{equation}
where second order terms like ${\kappa}^{2}$ have been dropped. The PDEs can be solved to yields the
following solutions
\begin{equation}
B_{s}=J_{0}(\sqrt{\frac{{\gamma}_{1}}{\eta}}r)e^{-{\gamma}_{1}t}
\label{25}
\end{equation}
This toroidal field would decay in time. When ${\gamma}_{1}=\gamma$ vanishes, due to the presence of
the Bessel function $J_{0}$, the toroidal field oscillates radially. As consequence, the
magnetic field does not grow in time, but yet could be called a marginal dynamo or
non-dynamo. The poloidal component $B_{\theta}$ can be easily computed in the case the second-
variations of $B_{\theta}$ decays very fast. This reduces its PDE above to
\begin{equation}
[{{\partial}_{r}}-[\frac{{\gamma}r}{\eta}+\frac{1}{r}]]B_{\theta}=0
\label{26}
\end{equation}
making the approximations $\frac{{\gamma}r}{\eta}<<\frac{1}{r}$ and
$r\approx{0}$ to be valid close to the magnetic axis, this
equation is reduces to
\begin{equation}
[{{\partial}_{r}}-\frac{1}{r}]B_{\theta}=0
\label{27}
\end{equation}
which immeadiatly yields $B_{s}=B_{0}re^{-{\gamma}_{1}t}=B_{0}r$ in the marginal dynamo case ${\gamma}_{1}=0$.
\section{Riemannian untwisted flux tube slow dynamos in
diffusive media}
In the first part of this section, it is shown that there is a relation between Riemann
curvature, which justifies to call these tubes with Frenet curved untwisted and
untorsioned Riemannian tubes. The Riemann curvature $R_{ijkl}$ where, (i,j = 1,2,3),
can be expressed as
\begin{equation}
R_{ijkl} = {\partial}_{[i}{\Gamma}_{j]kl} + {{\Gamma}^{p}}_{i[j}{\Gamma}_{p\slash{j}]}
\label{28}
\end{equation}
where ${\Gamma}_{ijk}$ is the Riemann-Christoffel symbols which can be expressed in terms of
the Riemann metric components $g_{ij}$ of the line element
\begin{equation}
ds^{2}= g_{ij}dx^{i}dx^{j} \label{29}
\end{equation}
by
\begin{equation}
{\Gamma}_{ijk}= \frac{1}{2}[g_{ij,k}+g_{ik,j}-g_{ij,k}]
\label{30}
\end{equation}
Here, the line element of the untwisted thick flux tube above one obtains
\begin{equation}
R_{1313}= R_{rsrs}= -\frac{K^{4}}{2r^{2}}=-\frac{1}{2}r^{2}{\kappa}^{4}cos^{2}{\theta}
\label{31}
\end{equation}
\begin{equation}
R_{2323}= -{K^{2}}
\label{32}
\end{equation}
This shows that the arguement above is valid. Back to the
self-induction equation,for thick tubes in diffusive medium, one notes that the only
difference now resides on the first term of the equation (\ref{10}). Due to thickness
hypothesis this leads to the extra equation
\begin{equation}
\frac{v_{s}}{rcos{\theta}}= {\eta}\frac{d}{ds}\kappa
\label{33}
\end{equation}
If one assumes, as before, that the curvature is constant, the dynamo action is not
present. Nevertheless, if one assumes a more general non-vanishing curvature, the
solution can be obtained as follows: Since by assumption the toroidal flow depends
only on the radial coordinate-r, the only way out, equation (\ref{33}) is to impose the
following constraint
\begin{equation}
{v_{s}}= {\eta}r
\label{34}
\end{equation}
\begin{equation}
\frac{d}{ds}\kappa=\frac{1}{cos{\theta}}
\label{35}
\end{equation}
Expression (\ref{34}) shows that the Riemannian °ow is shear °ow, since the °ow is
toroidal and depends linearly on another direction. As long as the tube is untwisted,
the trigonometric function does not depend upon the toroidal coordinate-s. This in
turn, yields the solution of equation (\ref{34}) as
\begin{equation}
\kappa=\frac{s}{cos{\theta}}+c_{0}
\label{36}
\end{equation}

From this expression, one is able to compute the Riemann curvature components
for this slow dynamo flow as
\begin{equation}
R_{1313}= R_{rsrs}= -\frac{1}{2}r^{2}{s}^{4}cos^{6}{\theta}
\label{37}
\end{equation}
\begin{equation}
R_{2323}= -{r^{2}}
\label{38}
\end{equation}
Note that component $R_{2323}$ grows very fast, when one walks along the magnetic
flux tube axis toroidal direction.
\section{Magnetic diffusive filaments with normal stretching and dynamo action}
Let us define the poloidal flow $v_{P}$ in terms of Frenet frame of the magnetic filament
$(t,n,b)$ as
\begin{equation}
\textbf{v}_{P}= v_{n}\textbf{n} + v_{b}\textbf{b}\label{39}
\end{equation}
and the toroidal flow as
\begin{equation}
\textbf{v}_{T}= v_{s}\textbf{t}\label{40}
\end{equation}
Substitution of these expressions into the diffusion-free self-induction equation
\begin{equation}
d_{t}\textbf{B}= \nabla\times(\textbf{v}\times\textbf{B})
\label{41}
\end{equation}
where the poloidal and toroidal components of the magnetic field can be expressed
in analogy to the components of the flow, results in the following equation
\begin{equation}
{\gamma}B_{s}=0\rightarrow{{\gamma}=0}
\label{42}
\end{equation}

when a toroidal component of the magnetic field is present. This implies that the
remaining equation is
\begin{equation}
{v_{s}}\tau (B_{n} + B_{b})=0\label{43}
\end{equation}
In the absence of toroidal flow, ${v_{s}}=0$ yields
\begin{equation}
B_{n}= -B_{b}\label{44}
\end{equation}
which considers also that the torsion is present in the filaments. In the absence of
toroidal component of the magnetic field $B_{s}=0$ it is clear that dynamo action is
possible, since equation (\ref{42}) does not necessarily implies that vanishes. This
is called a marginal dynamo \ref{21} in the dynamo theory literature. This case yields
\begin{equation}
\gamma={v_{s}}\tau\label{45}
\end{equation}
which shows that only poloidal flow $(v_{s}=0)$ cannot drive dynamo action as in Love
Gubbins case. Before introducing diffusion, one must note that the presence of
normal flows $v_{n}$ on filaments leads necessarilly to the compressibility of the flow.
This can be easily shown by considering the computation of the divergence of the
flow as
\begin{equation}
div\textbf{v}=v_{n}\kappa\label{46}
\end{equation}
which is distinct from zero for curved filaments such appears in the turbulent regime.
In the diffusive case one has
\begin{equation}
div\textbf{v}=v_{n}\kappa\label{47}
\end{equation}
where one has considered a helical filament here the torsion coincides with the
curvature and are constants. The growth rate of magnetic fields are given by
\begin{equation}
\gamma={\tau}_{0}[1+\eta{\tau}_{0}]\label{48}
\end{equation}
Since ${\tau}_{0}={\kappa}_{0}$, this expression shows that in the diffusion-free limit, the fast
dynamo action ${\gamma} > 0$ is driven by curves on negative scalar curvatures $({\kappa}_{0} < 0)$, in
strong analogy to the Chicone et al discover that the geodesic of negative curvature
compact Riemannian manifolds may drive fast dynamo action.
\section{Conclusions}
Anti-dynamo theorems, such as Cowling's can be bypass in certain physical situations in turbulent flows. In this paper it is shown that the Cowling's theorem is not violated when the flux
tube is thin and untwisting, whose curvature is Riemann-flat in diffusionless case. Riemann
metric curvature of thick flux tubes is shown to be related to the Frenet curvature. Cowling's
theorem is shown to be bypassed in a sort of violation in the diffusive case. Actually,
when the tube is thin, and the medium is diffusive the Cowling's theorem is not
applied even if the tube is axisymmetric as the example of Gissinger et al. The
dynamo action is however slow in this case, and a marginal dynamo is obtained. In
this case the tube is also Riemann-flat. In a Riemannian general case the Riemann
curvature does not vanish for a slow dynamo action as well for an untwisting thick
tube. At the end one must conclude that the absence of twist along with the presence of stretching and folding, nay be enough to get a slow dynamo action in the
untwisted flux tube. Analogy between the geodesic fast dynamo action in compact
constant negative curvature Riemannian manifolds and the Frenet negative constant
curvature slow dynamo, in magnetic curved filaments in turbulent regime.
\section{Acknowledgements}
I am deeply greateful to Andrew Soward, Jean Luc Thiffeault, Andrew D. Gilbert and Renzo
Ricca for their extremely kind attention and discussions on the subject of this paper. Thanks
are also due to I thank financial supports from Universidade do Estado do Rio de Janeiro
(UERJ) and CNPq (Brazilian Ministry of Science and Technology).

\end{document}